\documentclass{ws-ijmpcs}

\begin{document}

\markboth{Aubin \& {\it et al.}}
{Muon anomalous magnetic moment from the lattice}

%
\catchline{}{}{}{}{}
%

\title{THE MUON ANOMALOUS MAGNETIC MOMENT,\\ A VIEW FROM THE
LATTICE
}
\author{CHRISTOPHER AUBIN}

\address{Department of Physics and Engineering Physics, Fordham University\\
Bronx,
NY 10458, USA}

\author{THOMAS BLUM}

\address{Physics Department, 
University of Connecticut\\
Storrs, CT 06269, USA}

\author{MAARTEN GOLTERMAN\footnote{
Speaker at conference}}

\address{Department of Physics and Astronomy, San Francisco State University\\
San Francisco, CA 94132, USA}

\author{KIM MALTMAN}

\address{Department of Mathematics and Statistics,
York University\\
Toronto, ON Canada M3J~1P3}

\author{SANTIAGO PERIS}

\address{Department of Physics, Universitat Aut\`onoma de Barcelona\\
E-08193 Bellaterra, Barcelona, Spain}

\maketitle

\begin{history}
\received{Day Month Year}
\revised{Day Month Year}
\end{history}

\begin{abstract}
We review some of the issues that arise in attempts to compute the
hadronic corrections to the muon anomalous magnetic moment using
Lattice QCD.  We concentrate on the dominant contribution, which 
requires an accurate evaluation of the hadronic vacuum polarization. 
\end{abstract}
\vskip0.7cm

It is well-known that at present there exists a discrepancy of about
3 to 3.5 $\sigma$ between the experimental value for the
muon anomalous magnetic moment, $a_\mu=\frac{1}{2}(g-2)_\mu$, and the
best theoretical estimate of its value currently available.\cite{reviews}
Since it is expected that the E989 experiment at Fermilab will improve
the experimental error by a factor 4, to about 0.15 ppm, it is important
that theory keeps up, both in bringing down the error on the theory
side, as well as by corroborating the reliability of the theory error.

The dominant source of error comes from the hadronic contributions to $a_\mu$, in particular from the leading-order hadronic vacuum polarization
correction (HVP), and from the hadronic light-by-light correction (HLxL).   It is therefore
natural to see whether Lattice QCD can provide first-principle computations
of these contributions with competitive errors.   This would be the first
completely theoretical computation of these quantities, since the current
best estimate for HVP is based on a dispersive analysis of the 
experimentally measured $e^+e^-\to\mbox{hadrons}$ cross section, while
all estimates for HLxL are based on models.   In order to be interesting,
such computations should reach at least an accuracy of order 1\% for
HVP, and 10-20\% for HLxL.   Here we will therefore focus on HVP.\cite{latrev}
One may also hope that lattice estimates might shed light on 
potential discrepancies between various non-lattice theory estimates, such as
the discrepancy between the $e^+e^-$ determination of 
HVP, and the estimate in which the $I=1$ part of the spectral function
is obtained from $\tau$ decays.\footnote{For a recent discussion, and
references, see Ref.~\refcite{JS2012}.}

Eventually, it will be possible to put quarks, gluons and photons on the
lattice, in a combined Lattice QCD$+$QED framework.   This system
can then be probed with a muon, obtaining the full hadronic contribution to
$a_\mu$.   This approach is still in the future, although attempts to use
this approach are being explored for HLxL.\cite{latrev}   At present,
significantly smaller errors for HVP can be expected by treating the
photons as a perturbation, using the lattice to compute the purely
hadronic vacuum polarization only.\footnote{This makes a direct
comparison with the non-lattice theory value for HVP non-trivial,
since the $e^+e^-\to\mbox{hadrons}$ cross section contains higher-order
corrections in $\alpha$.}   

In the latter approach\cite{amuHVP}
\begin{equation}
\label{amu}
a_\mu^{\rm HVP}=4\alpha^2\int_0^\infty dQ^2\,f(Q^2,m_\mu^2)
\left(\Pi(0)-\Pi(Q^2)\right)\ ,
\end{equation}
where $\Pi(Q^2)$ as a function of euclidean $Q^2$ is obtained from the hadronic vacuum polarization
$\Pi_{\mu\nu}(Q)=(Q^2\delta_{\mu\nu}-Q_\mu Q_\nu)\Pi(Q^2)$,
$f(Q^2,m_\mu^2)$ is a known kinematical weight function, and $\alpha$ is the fine-structure
constant.

\begin{figure}[t]
\centerline{\includegraphics[width=7cm]{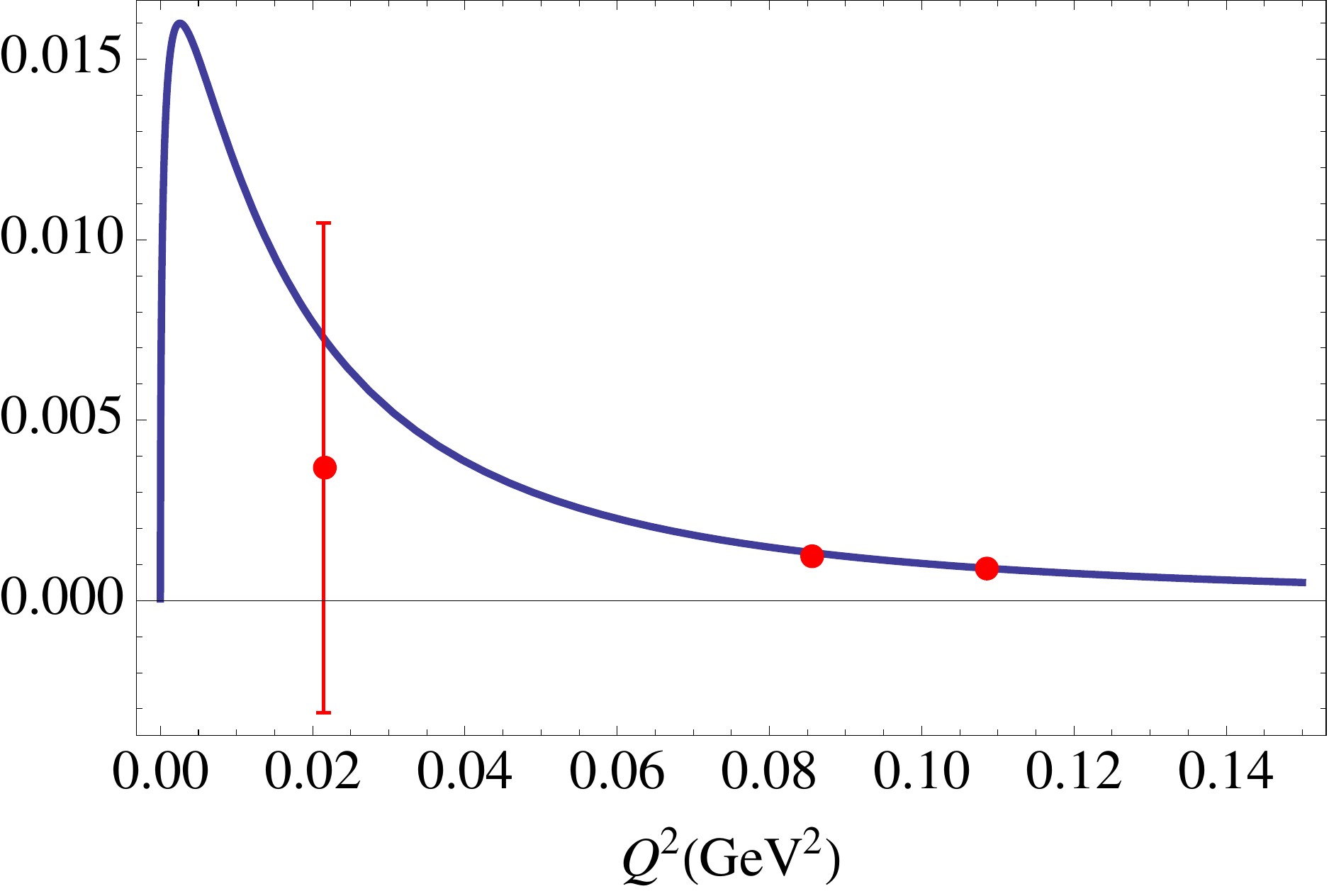}}
\vspace*{8pt}
\caption{Low-$Q^2$ behavior of the integrand 
$ f(Q^2,m_\mu^2)(\Pi(0)-\Pi(Q^2))$ in Eq.~(\ref{amu}). 
Red points show typical data on a $64^3\times 144$
lattice with lattice spacing $0.06$~fm and periodic boundary conditions.\label{f1}}
\end{figure}

Fig.~\ref{f1} shows the integrand of Eq.~(\ref{amu}) in the low-$Q^2$ region;
it is peaked around $Q^2\sim m_\mu^2/4$.
The points with error bars show typical lattice data for this integrand.\cite{ABGP2012}  The 
lattice data are restricted to  values of $Q^2$ available with periodic 
boundary conditions in finite volume.
The figure demonstrates an important problem in obtaining HVP from the
lattice:  to obtain lattice data near the peak of the integrand,
lattices with a linear volume $2\pi /L\sim m_\mu/2$, {\it i.e.}, $L\sim 25$~fm would be needed!   With a lattice spacing $a$ of order $0.06$~fm necessary 
in order  
to reach the continuum limit, this would imply $L/a\sim 400$, clearly out
of reach given present computational resources.

There are two ways in which one might proceed in order to make progress.
One method is clearly to obtain more data at smaller $Q^2$.   If one makes
use of twisted instead of periodic boundary conditions, one can in 
principle reach smaller $Q^2$ values without increasing the size of the
lattice.   Investigations in this direction are in progress.\cite{twisted}
The other is to obtain more precise data at currently available values of
$Q^2$, using for instance AMA error reduction.\cite{BIS}   If a theoretically reliable
fit function for the $Q^2$ behavior of $\Pi(Q^2)$ can be found, it may then be
possible to extrapolate the integrand of Eq.~(\ref{amu}) to smaller
values of $Q^2$, so that the integral $a_\mu^{\rm HVP}$ can be
computed with a small enough error.   Quite likely, a combination of these
methods will be necessary in practice.

The most commonly used fitting functions are based on the assumption of
vector meson dominance (VMD).\cite{vmd}  The problem with these
is that such fits assume that the lowest singularity in $\Pi(Q^2)$ is 
at $-Q^2=m_\rho^2$, while in reality $\Pi(Q^2)$ has a cut starting at $-Q^2=4m_\pi^2
\ll m_\rho^2$.   Clearly, the use of this assumption introduces a
model element into the computation, in conflict with the notion of
the lattice providing us with an approach from first principles!

Theoretically, one can do much better.   Based on results obtained in the
literature on Pad\'e approximants (PAs), it was proven in Ref.~\refcite{ABGP2012} that the functions
\begin{equation}
\label{PA}
\Pi(Q^2)=\Pi(0)-Q^2\left(a_0+\sum_{n=1}^{[P/2]}\frac{a_n}{b_n+Q^2}
\right)\ ,
\quad a_{n\ge 1}>0\ ,\quad b_n\ge 4m_\pi^2
\end{equation}
with either $a_0=0$ or $a_0$ free provide a series of PAs converging to the vacuum polarization everywhere
except near the cut $Q^2\in(-\infty,-4m_\pi^2]$ on the Minkowski axis.
We note that choosing $P=2$, $a_0=0$ and $b_1=m_\rho^2$ corresponds to a VMD-type assumption,
but it does not correspond to a valid PA:  as we increase the order of the
PA, the poles in Eq.~(\ref{PA}) should approach the branch point at $Q^2=-4m_\pi^2$.

While initial explorations of the PA-based fitting method of the low-$Q^2$ 
behavior of $\Pi(Q^2)$ look promising,\cite{ABGP2012} it is important to
have an independent test of any fitting method.   This is particularly 
important as long as the data in the strongly peaked region of Fig.~\ref{f1}
will remain sparse.   
Reference~\refcite{GMP2013} describes the construction of a
QCD-based model that allows us to set up a 
``test laboratory'' for fits of the low-$Q^2$ behavior of $\Pi(Q^2)$.\footnote{The
$Q^2>1$~GeV$^2$ part of Eq.~(\ref{amu}) can reliably be obtained
from a trapezoidal rule approximation of the integral.\cite{GMP2013}}
This model combines the non-strange vector $\tau$ spectral data with a
quantitative description beyond the $\tau$ mass using perturbation theory
and a model for duality violations in order to create a model for $\Pi(Q^2)$
using a dispersion relation.   In this model, the ``exact'' value of the integral
below $1$~GeV$^2$ is
\begin{equation}
\label{exact}
\tilde{a}_\mu^{\mbox{HVP},Q^2\le 1~\mbox{GeV}^2}=1.204\times 10^{-7}\ ,
\end{equation}
where the tilde reminds us that this quantity is not $a_\mu$ itself,
but instead corresponds to twice the $I=1$ part below 1~GeV$^2$. 
Using a covariance matrix for $\Pi(Q^2)$ data points obtained from a 
lattice computation, we may now use the model in order to create 
realistically correlated fake data
sets at typical lattice values of the momenta $Q^2$.   These data sets
can then be fitted using VMD- or PA-type fit functions, allowing us to compare
fitted values for $\tilde{a}_\mu^{\mbox{HVP},Q^2\le 1~\mbox{GeV}^2}$
with the exact value in Eq.~(\ref{exact}).

\begin{table}[t]
\tbl{Various correlated fits of the ``lattice'' data set
on the interval $0<Q^2\le 1$ GeV$^2$.}
{\begin{tabular}{@{}ccccc@{}} \toprule
Fit & $\tilde{a}_\mu\times 10^7$ & error $\times 10^7$ &
pull & $\chi^2/\mbox{dof}$ \\
\colrule
VMD & 1.3201 & 0.0052 & 22 & 2189/47 \\
VMD$+$ & 1.0658 & 0.0076 & 18 & 67.4/46 \\
\colrule
PA $[0,1]$ & 0.8703 & 0.0095 & 35 & 285/46 \\
PA $[1,1]$ & 1.116 & 0.022 & 4 & 61.4/45 \\
PA $[1,2]$ & 1.182 & 0.043 & 0.5 & 55.0/44 \\
PA $[2,2]$ & 1.177 & 0.058 & 0.5 & 54.6/43 \\ \botrule
\end{tabular} \label{t1}}
\end{table}

Results for such fits are shown in Table~\ref{t1}.   VMD$+$ corresponds
to VMD plus a linear term, and the PA fits correspond to Eq.~(\ref{PA})
with $P\ge 2$, $a_0=0$ or $a_0$ free.   For details, we refer to Ref.~\refcite{GMP2013}.   The first column lists the type of fit, the second
column gives the value of $\tilde{a}_\mu^{\mbox{HVP},Q^2\le 1~\mbox{GeV}^2}$ from the fit, the third column the error from the fit, and
the fifth column the value of $\chi^2$ per degree of freedom for the fit.   
From the
$\chi^2$ values, one would conclude the VMD$+$ fit to be reasonable,
and one would accept the PA fits $[1,1]$, $[1,2]$ and $[2,2]$, if no extra information
were available.

\begin{figure}[t]
\centering
\includegraphics[width=2in]{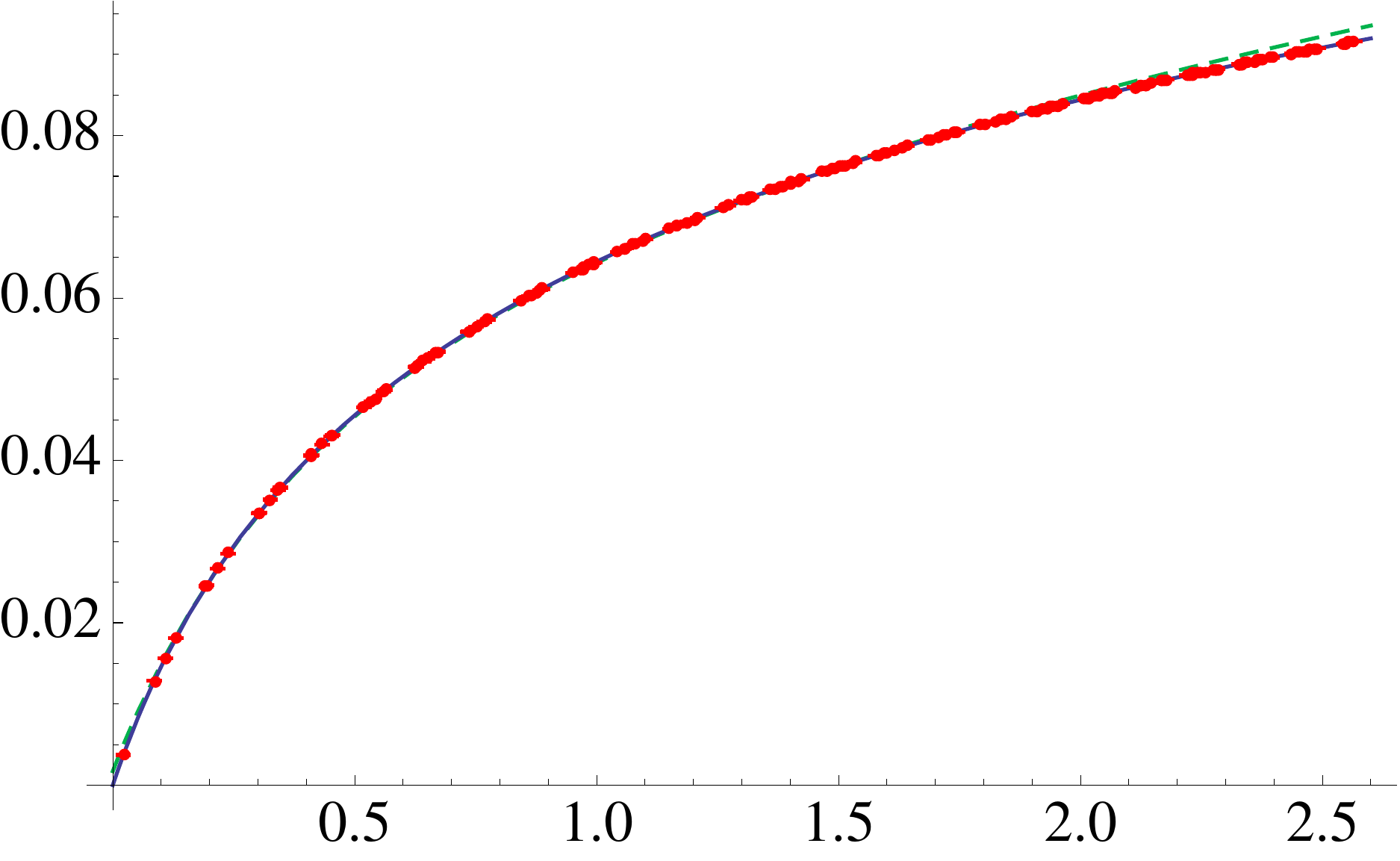}
\hspace{.4cm}
\includegraphics[width=2.2in]{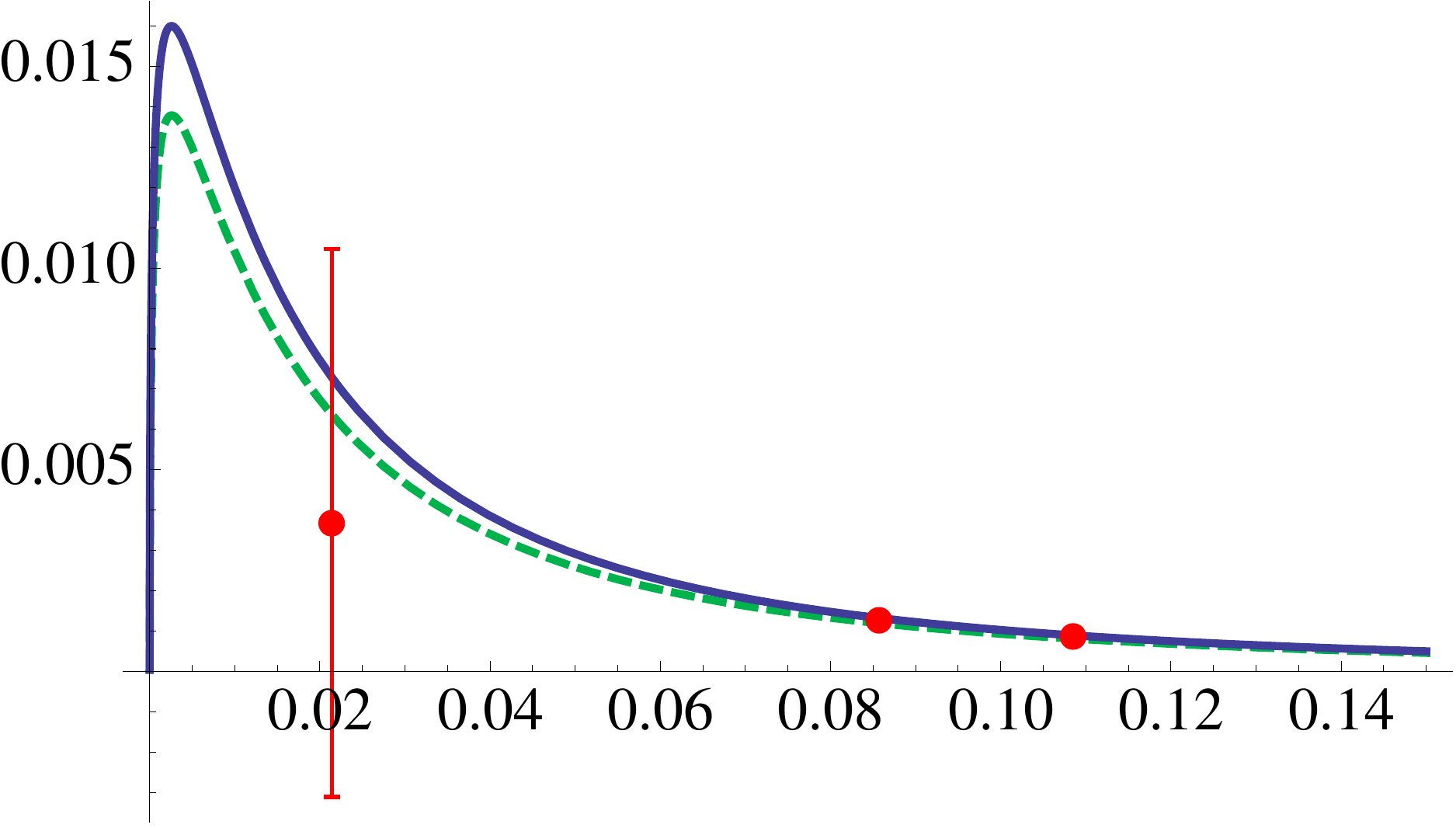}
\vspace*{8pt}
\caption{VMD$+$ fit: green dashed curves;  model: blue solid curves. Left panel: vacuum polarization; right panel: 
integrand of Eq.~(\ref{amu}) for low $Q^2$; red points are
``lattice'' data points. Units as in Fig.~\ref{f1}.  \label{f2}}
\vspace*{-13pt}
\end{figure}

However, in our model tests, we can compare the fitted value with the
exact value.   The fourth column shows the ``pull,'' defined by
\begin{equation}
\label{pull}
\mbox{pull}=\frac{|\mbox{exact~value}-\mbox{fitted~value}|}{\mbox{fit~error}}\ .
\end{equation}
From its values, we see that both VMD-type fits are very bad fits;
there is a large discrepancy between the precision as measured by the
fit error and the accuracy as measured by the pull (even though
the VMD$+$ fit has an acceptable $\chi^2/$dof).   Also the $[1,1]$ PA
is not very good --- only the higher order PAs yield good fits, but the 
fit error is of order 5\%.  For illustration, we show the VMD$+$ fit in 
Fig.~\ref{f2}.   While the $\Pi(Q^2)$ fit looks very good, this is not the
case for the integrand of Eq.~(\ref{amu}).   A similar figure for the $[1,2]$
PA looks much better.\cite{GMP2013}

In conclusion, we believe that it is important to pursue $a_\mu^{\rm HVP}$ using Lattice QCD, but it is clear that much work
remains in order to ascertain the accuracy with which this can be done.
In particular, we advocate the use of fit functions for the low-$Q^2$ 
behavior of $\Pi(Q^2)$ based on known convergence properties, and
the use of detailed tests as sketched above in order to test fit strategies
as part of the error analysis of any determination of $a_\mu^{\rm HVP}$
from the lattice.

\vspace*{-5pt}
\section*{Acknowledgments}

TB and MG are supported by the US DOE under
Grants DE-FG02-92ER40716 and DE-FG03-92ER40711,
KM  by  NSERC (Canada), and
SP  by the  Mininisterio de Educaci\'on (Spain) under Grants CICYTFEDER-FPA2011-25948, SGR2009-894.

\end{document}